\documentclass[sigconf]{acmart}

\usepackage{lmodern}
\usepackage{hyperref}
\usepackage{tabularx}
\usepackage{booktabs}
\usepackage{url}
\usepackage{xcolor}
\usepackage{caption}
\usepackage{enumitem}
\usepackage{tikz}
\usetikzlibrary{
  patterns,
  positioning,
  arrows.meta,
  shapes.geometric,
  shapes.symbols,
  fit,
  calc,
  backgrounds
}

\definecolor{SpecOrange}{RGB}{245,166,35}
\definecolor{KnownBlue}{RGB}{170,220,255}
\definecolor{SpecBlue}{RGB}{50,110,255}
\definecolor{UnknownRed}{RGB}{255,140,140}
\definecolor{ToolsPurple}{RGB}{190,175,255}
\definecolor{ToolYellow}{RGB}{245,235,170}
\definecolor{ToolGreen}{RGB}{40,160,70}

\title{Securing the Agent: Vendor-Neutral, Multitenant Enterprise Retrieval and Tool Use}

\author{Francisco Javier Arceo}
\affiliation{%
  \institution{Red Hat AI}
  \city{Boston}
  \country{USA}
}
\email{farceo@redhat.com}

\author{Varsha Prasad Narsing}
\affiliation{%
  \institution{Red Hat AI}
  \city{Boston}
  \country{USA}
}
\email{vnarsing@redhat.com}

\begin{abstract}
Retrieval-Augmented Generation (RAG) and agentic AI systems are increasingly prevalent in enterprise AI deployments. However, real enterprise environments introduce challenges largely absent from academic treatments and consumer-facing APIs: multiple tenants with heterogeneous data, strict access-control requirements, regulatory compliance, and cost pressures that demand shared infrastructure.

A fundamental problem underlies existing RAG architectures in these settings: retrieval systems rank documents by relevance---whether through semantic similarity, keyword matching, or hybrid approaches---not by authorization, so a query from one tenant can surface another tenant's confidential data simply because it scores highest. We formalize this gap and analyze additional shortcomings---including tool-mediated disclosure, context accumulation across turns, and client-side orchestration bypass---that arise when agentic systems conflate relevance with authorization. To address these challenges, we introduce a layered isolation architecture combining policy-aware ingestion, retrieval-time gating, and shared inference, enforced through server-side agentic orchestration. This approach centralizes security-critical operations---tool execution authorization, state isolation, and policy enforcement---on the server, creating natural enforcement points for multitenant isolation while allowing client-side frameworks to retain control over agent composition and latency-sensitive operations.

We validate the proposed architecture through an open-source implementation in OGX\footnote{OGX (Open GenAI Stack), formerly known as Llama Stack~\cite{llamastack}. The project is available at {\color{blue}\url{https://github.com/ogx-ai/ogx}}; the Kubernetes Operator at {\color{blue}\url{https://github.com/ogx-ai/ogx-k8s-operator}}.}~\cite{ogx}, a vendor-neutral framework that implements an OpenAI-compatible, open-source Responses API with server-side multi-turn orchestration. We evaluate it empirically and show that ABAC gating eliminates cross-tenant leakage while introducing negligible overhead.
\end{abstract}

\ccsdesc[300]{Computing methodologies~Machine learning}
\ccsdesc[100]{Information systems~Data management systems}

\keywords{multitenancy, retrieval-augmented generation, RAG, access control, agentic AI, server-side orchestration, open source, OGX, LLM systems, LLMOps}

\begin{document}

\acmYear{2026}\copyrightyear{2026}
\acmConference[ACM CAIS '26]{ACM Conference on AI and Agentic Systems}{May 26--29, 2026}{San Jose, CA, USA}
\acmBooktitle{ACM Conference on AI and Agentic Systems (ACM CAIS '26), May 26--29, 2026, San Jose, CA, USA}
\acmDOI{10.1145/3786335.3813145}
\acmISBN{979-8-4007-2415-2/26/05}

\maketitle

\section{Introduction}

Enterprise adoption of agentic AI introduces security challenges that existing architectures do not address. We motivate the problem, formalize it, and summarize our contributions.

\subsection{Motivation}
Enterprise adoption of generative AI has evolved beyond simple prompt-response interactions toward agentic systems---AI applications that autonomously reason, use tools, retrieve information, and execute multi-step workflows to accomplish complex tasks~\cite{yao2023react,schick2023toolformer,karpas2022mrkl}.
This evolution reflects a fundamental shift: rather than treating large language models (LLMs) as sophisticated text generators, organizations now deploy them as reasoning engines capable of taking actions in the world.

An API-first paradigm is emerging to unify multi-turn inference, tool use, and retrieval behind a single endpoint.
OpenAI's Responses API is one prominent example, providing a unified interface for chat-style inference, tool invocation, retrieval, and stateful workflows~\cite{openaiResponsesAPI}, while open-source frameworks increasingly expose OpenAI-compatible endpoints to decouple applications from any single provider.

However, real-world enterprise deployments differ sharply from the assumptions embedded in consumer-facing APIs and many academic prototypes, exhibiting characteristics that demand specialized architectural consideration:
\begin{itemize}
  \item \textbf{Multiple tenants:} distinct business units, customers, or partners served from shared infrastructure, with strict isolation requirements.
  \item \textbf{Heterogeneous data:} document collections vary in format, sensitivity classification, and access requirements.
  \item \textbf{Strict access control:} regulatory frameworks require fine-grained governance with auditable access patterns.
  \item \textbf{Operational control:} visibility into agent behavior, tool execution sequences, and data access patterns is required for debugging and compliance, consistent with lessons from production ML engineering~\cite{amershi2019se4ml}.
  \item \textbf{Vendor independence:} lock-in to a single AI provider creates business risk; enterprises require on-prem and hybrid options.
\end{itemize}

Na\"{\i}ve approaches to addressing these requirements replicate the entire agentic stack per tenant: separate vector stores, dedicated inference endpoints, and isolated tool configurations.
This strategy incurs substantial costs---infrastructure scales linearly with tenants rather than with actual usage---and creates operational fragmentation that amplifies common ML systems maintenance risks~\cite{sculley2015hiddentechnicaldebt}.

\subsection{Problem statement}
This paper addresses a fundamental tension in enterprise agentic AI deployment:

\begin{quote}
Autonomous agents require flexible tool access and multi-turn reasoning capabilities, yet enterprise environments demand strict tenant isolation and policy enforcement---requirements that existing agentic architectures cannot simultaneously satisfy.
\end{quote}

Standard agentic AI deployments exhibit security assumptions incompatible with enterprise multitenancy:
\begin{enumerate}
  \item \textbf{Client-side orchestration:} the application manages the inference $\rightarrow$ tool $\rightarrow$ inference loop, distributing security-critical logic to potentially untrusted clients and increasing operational complexity~\cite{amershi2019se4ml}.
  \item \textbf{Homogeneous data access:} retrieval stacks assume uniform access to a corpus; retrieval methods (whether dense, sparse, or hybrid) optimize relevance ranking rather than authorization~\cite{karpukhin2020dpr}.
  \item \textbf{Implicit trust boundaries:} tool execution is often treated as a capability extension without systematic verification of who may invoke tools or consume tool outputs, despite the centrality of tool use in modern agent designs~\cite{yao2023react,schick2023toolformer,karpas2022mrkl}.
  \item \textbf{Stateless isolation:} requests are treated independently, ignoring how conversation state and cached tool results can leak across boundaries; such hidden couplings are a classic source of ML systems fragility~\cite{sculley2015hiddentechnicaldebt}.
\end{enumerate}

In multitenant settings, these assumptions create serious vulnerabilities.
A document highly similar to a query may belong to a different tenant.
A tool call may access resources outside the user's authorization scope.
Conversation history may accumulate context that crosses security boundaries.

This paper focuses on agentic RAG---the intersection of retrieval-augmented generation and autonomous tool use---as the primary case study, since retrieval is where the relevance-authorization gap is most acute. However, the layered isolation architecture and server-side enforcement patterns we propose are general and extend to other agentic capabilities including inference, tool execution, code generation, and multi-step workflows.

\subsection{Contributions}
This paper makes the following contributions:
\begin{enumerate}
  \item We formalize the \emph{relevance-authorization gap} in multitenant RAG, showing why relevance ranking alone is insufficient to enforce isolation without explicit authorization predicates~\cite{karpukhin2020dpr,johnson2017faiss}, and provide empirical evidence that ungated retrieval leaks cross-tenant data in 98--100\% of probes.

  \item We propose a layered isolation architecture combining policy-aware ingestion, retrieval-time gating, and shared inference, with server-side orchestration as the enforcement layer for the agentic control loop.

  \item We validate the architecture through a six-experiment evaluation measuring security (cross-tenant leakage, prompt injection resilience), systems performance (latency overhead, throughput scaling), and retrieval quality across a 2$\times$2 configuration matrix crossing orchestration mode with retrieval gating.

  \item We implement the architecture in OGX~\cite{ogx}, an open-source, vendor-neutral framework providing OpenAI-compatible APIs with pluggable providers for inference, vector stores, and tools, deployable on Kubernetes via a dedicated operator~\cite{ogxk8soperator}.
\end{enumerate}

\section{Background}

We trace the evolution of LLM application architectures to show how each phase introduced new capabilities but also inherited the security gaps of prior phases.

\subsection{The evolution of LLM application architectures}
LLM application architectures have evolved through distinct phases, each introducing new capabilities and security considerations.
The \textbf{Completions APIs} exposed simple prompt\,$\rightarrow$\,text interfaces with perimeter-oriented security.

\textbf{Retrieval-augmented generation (RAG)}~\cite{lewis2020rag} introduced retrieval and new attack surfaces such as retrieval manipulation and context poisoning, building on foundational work on prompt injection vulnerabilities~\cite{willison2022prompt}.
\textbf{Dense retrieval and learned retrievers} (e.g., DPR~\cite{karpukhin2020dpr} and retrieval-augmented pretraining such as REALM~\cite{guu2020realm}) established the core technical basis for modern RAG.
\textbf{Retrieval infrastructure} evolved to support multiple search modalities: dense vector search across diverse vector databases~\cite{johnson2017faiss}, sparse keyword matching (BM25), and hybrid approaches combining both with neural rerankers~\cite{blendedrag2024}. However, all of these modalities optimize for relevance; none enforce authorization natively.
\textbf{Tool-using agents}~\cite{yao2023react,schick2023toolformer,karpas2022mrkl} extended LLMs with tool calls and an inference\,$\rightarrow$\,tool\,$\rightarrow$\,inference loop.
\textbf{Autonomous multi-step agents} execute multi-tool workflows with limited human oversight.
\textbf{The Responses API paradigm}~\cite{openaiResponsesAPI,openresponses} represents a convergence of these phases, unifying inference, tool use, retrieval, and state management behind a single endpoint.

Each phase introduced new capabilities but also inherited the security gaps of prior phases. The Responses API paradigm unifies these capabilities under a single interface but assumes single-tenant deployment. Enterprise multitenancy requires policy-aware tool execution, stateful conversation management, and orchestration controls---challenges amplified in agentic deployments where reasoning, retrieval, and tool execution interleave autonomously~\cite{sculley2015hiddentechnicaldebt,amershi2019se4ml}.

\section{Architecture}
\label{sec:architecture}

We formalize the multitenant agentic AI environment: tenants ($T$) share infrastructure, each with associated data, users, tools, and policies. Agent execution follows the tool-using pattern~\cite{yao2023react,karpas2022mrkl}, where an execution sequence $E$ consists of alternating inference steps $i$, tool calls $\phi$, and responses $r$:

\begin{equation}
E = [(i_1, \phi_1, r_1), (i_2, \phi_2, r_2), \dots, (i_n, \emptyset, r_n)]
\end{equation}

where the final step has no tool call ($\emptyset$) and terminates with response $r_n$.

Standard agentic deployments exhibit several security assumptions that are fundamentally incompatible with enterprise multitenancy.

\subsection{Security challenges}

\textbf{Relevance-authorization gap.} Retrieval systems---whether vector-based, keyword-based, or hybrid---optimize for relevance metrics rather than authorization policies. This creates a fundamental gap: search ranking considers semantic similarity, term frequency, or combined relevance signals, but authorization decisions depend on access control policies that are orthogonal to these relevance measures.

For tenants $T_A$ and $T_B$ sharing corpus $D = D_A \cup D_B$, retrieval methods cannot enforce tenant isolation without an authorization predicate. Let $q$ denote a query, $u$ denote a user, $d$ denote a document, $\theta$ denote a relevance threshold, and $P(u,d)$ denote an authorization policy that returns $\mathrm{permit}$ or $\mathrm{deny}$. Then secure retrieval requires:

\begin{equation}
\{ d \in D : \operatorname{relevance}\bigl(q,d\bigr) > \theta \wedge P\bigl(u,d\bigr) = \mathrm{permit} \}
\end{equation}

\textbf{Tool-mediated disclosure.} Agents invoke tools with agent credentials rather than end-user authorization, potentially accessing unauthorized data across tenant boundaries.

\textbf{Context accumulation.} Multi-turn conversations persist context without per-turn policy re-validation, enabling cross-tenant data leakage.

\textbf{Client-side bypass.} When orchestration runs client-side, malicious clients can skip authorization checks, manipulate tool invocation, or extract unauthorized data.

These shortcomings stem from distributing security-critical logic outside the trust boundary~\cite{sculley2015hiddentechnicaldebt,amershi2019se4ml,shapira2025agentsofchaos}.

\subsection{Threat model and assumptions}
\label{sec:threat-model}

We define the scope of adversaries and assumptions under which the architecture provides its security guarantees.

\textbf{In-scope adversaries:} (1)~a \emph{malicious tenant} crafting queries---including prompt injections---to retrieve another tenant's data; (2)~a \emph{compromised client} bypassing authorization by calling ungated endpoints or manipulating tool invocations; (3)~a \emph{buggy tool} leaking cross-tenant state through tool outputs or side effects.

\textbf{Out-of-scope adversaries:} a compromised server process, an insider operator with infrastructure access, side-channel attacks (timing, cache), and model extraction attacks. The trust boundary is the server process---all security-critical operations execute within it.

\textbf{System goals:}
\begin{itemize}
  \item[\textbf{G1}] No cross-tenant data leakage through retrieval.
  \item[\textbf{G2}] Authorization enforcement independent of orchestration mode.
  \item[\textbf{G3}] Denied queries fail fast without invoking inference.
\end{itemize}

\textbf{Assumptions:} (A1)~correct token-to-tenant mapping by the authentication provider; (A2)~immutable document ownership metadata assigned at ingestion; (A3)~the inference layer is untrusted---it may leak any context it receives, so isolation must be enforced before context construction; (A4)~vector backends faithfully apply metadata filters when predicate pushdown is supported.

\subsection{Data path: layered isolation}
\label{sec:data-path}
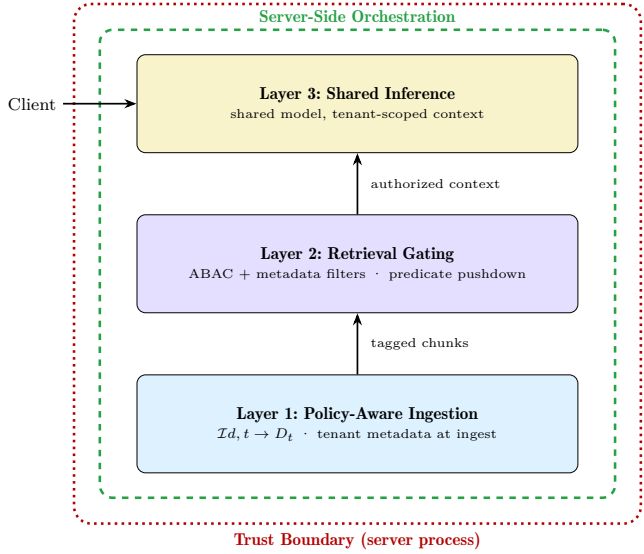
\begin{figure}[!htbp]
  \centering
  \resizebox{\columnwidth}{!}{%
  \begin{tikzpicture}[
    layer/.style={draw, rounded corners=4pt, minimum width=72mm, minimum height=16mm,
                  align=center, font=\small\bfseries},
    arrow/.style={-{Stealth[length=2mm]}, thick},
    every node/.style={inner sep=3pt}
  ]
    \node[layer, fill=KnownBlue!40] (L1) {Layer 1: Policy-Aware Ingestion\\[-1pt]{\scriptsize\normalfont $\mathcal{I}(d,t) \rightarrow D_t$ · tenant metadata at ingest}};
    \node[layer, fill=ToolsPurple!40, above=10mm of L1] (L2) {Layer 2: Retrieval Gating\\[-1pt]{\scriptsize\normalfont ABAC + metadata filters · predicate pushdown}};
    \node[layer, fill=ToolYellow!60, above=10mm of L2] (L3) {Layer 3: Shared Inference\\[-1pt]{\scriptsize\normalfont shared model, tenant-scoped context}};

    \draw[arrow] (L1) -- node[right, font=\scriptsize, xshift=1mm] {tagged chunks} (L2);
    \draw[arrow] (L2) -- node[right, font=\scriptsize, xshift=1mm] {authorized context} (L3);

    \node[draw=ToolGreen, very thick, rounded corners=6pt, dashed,
          fit=(L1)(L2)(L3),
          inner xsep=6mm, inner ysep=4mm,
          label={[font=\small\bfseries, text=ToolGreen]above:Server-Side Orchestration}] (orch) {};

    \begin{scope}[on background layer]
      \node[draw=red!70!black, very thick, dotted, rounded corners=8pt,
            fit=(orch),
            inner sep=4mm,
            label={[font=\small\bfseries, text=red!70!black]below:Trust Boundary (server process)}] {};
    \end{scope}

    \node[font=\small, left=12mm of L3] (client) {Client};
    \draw[arrow] (client) -- (L3.west);
  \end{tikzpicture}%
  }
  \caption{Layered isolation architecture with server-side orchestration. All three layers execute within the server trust boundary; the client controls \emph{what} to ask but not \emph{how} retrieval and tool execution are performed.}
  \label{fig:isolation-layers}
\end{figure}

To address the security challenges, we propose a two-part solution.
A three-layer isolation architecture secures the \emph{data path}: how documents are ingested, retrieved, and fed to the model (Figure~\ref{fig:isolation-layers}).
Server-side orchestration secures the \emph{control path}: how tools are invoked, state is managed across turns, and policies are enforced (Section~\ref{sec:control-path}).

\textbf{Layer 1: Policy-aware ingestion.} Tenant metadata attached at ingestion: $\mathcal{I}(d, t) \rightarrow D_t$ tags document $d$ with tenant $t$'s attributes. This ensures every chunk inherits ownership metadata, so downstream retrieval and authorization operate on consistent tenant attributes. Attaching metadata at ingestion rather than retrofitting it reduces the risk of accidental cross-tenant coupling~\cite{sculley2015hiddentechnicaldebt}.

\textbf{Layer 2: Retrieval gating.} Two-tier enforcement: resource-level ABAC authorization before search and chunk-level filtering after retrieval, composing similarity search with authorization predicates to implement Equation~(2). Where the backend supports predicate pushdown, tenant filters are applied natively during vector search, maintaining perfect recall regardless of corpus size. On backends without pushdown, post-retrieval filtering preserves the security guarantee (G1) but recall degrades at large corpus sizes as cross-tenant documents contaminate the top-$k$ set. We recommend pushdown-capable backends (e.g., pgvector, Qdrant, Milvus) for production deployments.

\textbf{Layer 3: Shared inference.} The LLM inference layer is shared across tenants; the model itself does not require per-tenant isolation, only the context fed to it. Because Layers~1 and~2 ensure that only authorized documents enter the prompt, the inference layer can be safely shared, reducing cost from $O$($N \cdot M$) to $O$($M$) where $N$ is the number of tenants and $M$ is the number of model endpoints.

This relies on assumption~A3: the architecture secures \emph{context construction}, not parametric memory. A model may generate information absorbed during pretraining regardless of retrieval gating. Per-tenant model instances can provide full parametric isolation but require deploying a separate model for each tenant, which is costly and impractical at scale. At the serving layer, modern systems show that batching, scheduling, and memory management dominate throughput and latency for generative transformers~\cite{yu2022orca,kwon2023vllm}.

\subsection{Control path: server-side orchestration}
\label{sec:control-path}

Client-side orchestration offers important advantages: low latency for local tool calls, flexibility in composing custom agent logic, and the ability to leverage rich client-side frameworks. However, in multitenant enterprise settings, purely client-side patterns expand the trusted computing base (TCB) to include untrusted client code. A compromised or buggy client can skip retrieval filters, invoke unauthorized tools, or accumulate cross-tenant context---the server cannot enforce security invariants it does not control~\cite{sculley2015hiddentechnicaldebt,amershi2019se4ml}.

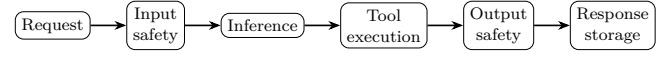
\begin{figure}[!htbp]
    \centering
    \resizebox{\columnwidth}{!}{%
    \begin{tikzpicture}[
      box/.style={draw, rounded corners, align=center, inner sep=3pt, font=\small},
      arrow/.style={-{Stealth[length=2.0mm]}, thick}
    ]
      \node[box] (req) {Request};
      \node[box, right=6mm of req] (in) {Input\\safety};
      \node[box, right=6mm of in] (inf) {Inference};
      \node[box, right=6mm of inf] (tool) {Tool\\execution};
      \node[box, right=6mm of tool] (out) {Output\\safety};
      \node[box, right=6mm of out] (store) {Response\\storage};
      \draw[arrow] (req) -- (in);
      \draw[arrow] (in) -- (inf);
      \draw[arrow] (inf) -- (tool);
      \draw[arrow] (tool) -- (out);
      \draw[arrow] (out) -- (store);
    \end{tikzpicture}%
    }
    \caption{Server-side orchestration flow: every step runs inside the server trust boundary.}
    \label{fig:orchestration-flow}
  \end{figure}

Server-side orchestration centralizes policy enforcement by executing the inference $\rightarrow$ tool $\rightarrow$ inference loop within the server trust boundary (Figure~\ref{fig:orchestration-flow}). This introduces trade-offs: added latency for tool calls that could execute locally and reduced flexibility for client-specific agent logic. For workloads where all tenants are trusted, client-side orchestration remains a pragmatic choice.

In practice, hybrid architectures are likely to emerge: server-side orchestration enforces security-critical invariants (authorization, state isolation, audit logging), while client-side frameworks retain control over agent composition and latency-sensitive operations.

Table~\ref{tab:enforcement} maps each security challenge to the enforcement point that mitigates it.

\begin{table}[!htbp]
  \centering
  \small
  \setlength{\tabcolsep}{3pt}
  \renewcommand{\arraystretch}{1.1}
  \begin{tabularx}{\columnwidth}{@{}l|X@{}}
  \toprule
  Challenge & Enforcement point \\
  \midrule
  Cross-tenant retrieval leakage & Layer 2 retrieval gating (ABAC + metadata filters) \\
  \midrule
  Context accumulation & Tenant-scoped state storage and per-turn authorization \\
  \midrule
  Tool-mediated disclosure & Server-side tool execution with authorization propagation \\
  \midrule
  Client-side bypass & Server-side orchestration (reduced TCB) \\
  \midrule
  Audit failure & Server-side telemetry and tracing \\
  \bottomrule
  \end{tabularx}
  \caption{Mapping from security challenges to architectural enforcement points.}
  \label{tab:enforcement}
\end{table}

\section{Implementation}
\label{sec:implementation}

OGX~\cite{ogx} provides an open-source implementation of the architecture proposed in Section~\ref{sec:architecture}. The framework exposes OpenAI-compatible APIs for inference and agentic execution, enabling any client-side framework (LangChain~\cite{langchain}, LangGraph~\cite{langgraph}, CrewAI~\cite{crewai}, and others) to gain server-side policy enforcement, multitenancy, and provider portability without changes to agent code. Together with open models such as gpt-oss~\cite{openai2025gptoss120bgptoss20bmodel}, this provides a complete open-source alternative to proprietary agentic AI platforms.

\subsection{APIs and agentic execution}
\label{sec:apis}

OGX defines over 20 APIs covering the full lifecycle of agentic applications. These span core capabilities---inference, agents (Responses API), vector stores and search, safety, tool runtime, and telemetry---as well as resource management APIs for models, files, file processors, prompts, conversations, connectors, and tool groups. The framework also provides compatibility layers for third-party API paradigms. Each API is designed with multitenancy as a first-class concern, enabling tenant-scoped resource management and access control. A complete API listing is provided in Appendix~\ref{sec:api-details}.

The Responses API---implementing the OpenAI Responses API paradigm~\cite{openaiResponsesAPI}---is the central orchestration endpoint (Figure~\ref{fig:openai-responses-api}). Unlike chat completion APIs that terminate after a single inference call, a single Responses API request may trigger multiple inference calls, tool executions, safety checks, and state transitions before producing a final response. All such operations execute within the server boundary: conversation state is retrieved and persisted server-side, tools are invoked under centralized authorization, and safety guardrails are applied at each step.

The Vector Stores and Search APIs provide uniform access to multiple retrieval modalities---dense vector search, keyword matching, and hybrid pipelines---with structured metadata filtering for tenant isolation. The Prompts API enables versioned prompt management, and the Conversations API maintains tenant-scoped multi-turn state. Together, these APIs provide the OpenAI-compatible interface through which the architecture's security guarantees are delivered: clients interact with standard endpoints (\texttt{/v1/responses}, \texttt{/v1/vector\_stores}, \texttt{/v1/chat/completions}) while the server enforces ABAC policies transparently.

\subsection{Provider architecture}
\label{sec:providers}

Extensibility is achieved through pluggable providers. Each API may be backed by multiple providers, categorized as \emph{inline} (executing within the OGX process) or \emph{remote} (adapting external services). This separation enables hybrid deployments: sensitive data paths remain local while computationally intensive operations are delegated to scalable external services. Crucially, provider substitution is transparent to clients---a developer can prototype with inline providers (e.g., sqlite-vec~\cite{sqlitevec}, an in-process safety model) and move to production-grade remote providers (e.g., pgvector, vLLM~\cite{kwon2023vllm}) without changing application code or security policies.

A routing layer dispatches API requests to provider instances based on logical resource identifiers, incorporating authorization checks and tenant identity before delegation. Different tenants may be routed to distinct provider instances while sharing the same API surface. A \emph{distribution} packages a specific set of APIs, providers, and resources into a deployable unit, decoupling application logic from provider selection. Supported providers are listed in Appendix~\ref{sec:api-details}.

\subsection{Access control}
\label{sec:access-control}

OGX includes a declarative, attribute-based access control (ABAC) framework that evaluates authorization at runtime. Access rules specify permit or deny scopes with conditions based on ownership and attribute matching. The default policy permits access when the user is the resource owner or when the user's attributes (roles, teams, projects, namespaces) match the resource's access attributes; a default-deny model ensures access is only granted when explicitly permitted.

Authorization is enforced at three levels: (1)~API route middleware, (2)~routing table resolution (before resolving a vector store or model to a provider), and (3)~storage read time, where query filters are constructed from the current user's attributes so that tenants only see their own or attribute-matched rows. JWT or Kubernetes authentication providers map external identity claims into these attributes, so enterprise identity systems drive isolation without embedding tenant IDs in application logic.

\subsection{Deployment}
\label{sec:deployment}

The OGX Kubernetes Operator~\cite{ogxk8soperator} automates deployment through custom resources that declaratively specify server configurations, backend connections, and isolation policies. The operator supports shared instances (multiple tenants with ABAC isolation), per-tenant instances (namespace-level isolation with Kubernetes RBAC), and hybrid approaches~\cite{burns2016borg}. In all topologies, the provider abstraction allows organizations to start with lightweight backends during development and migrate to production-grade infrastructure without modifying agent code or security policies.

\begin{figure}[!htbp]
  \centering
  \resizebox{\columnwidth}{!}{%
  \begin{tikzpicture}[
    box/.style={draw, rounded corners, align=center, inner sep=4pt, minimum height=8mm, font=\small},
    arrow/.style={-{Stealth[length=2.0mm]}, thick}
  ]
    \node[box] (client) {Client\\\texttt{POST /v1/responses}};
    \node[box, below=7mm of client] (auth) {AuthN/AuthZ\\tenant context};
    \node[box, below=7mm of auth] (orch) {Server-side orchestrator\\(multi-turn loop)};

    \node[box, below left=7mm and 12mm of orch] (infer) {Shared inference\\(LLM serving)};
    \node[box, below=7mm of orch] (vec) {Vector store\\(logical isolation)};
    \node[box, below right=7mm and 12mm of orch] (policy) {Policy engine\\(ABAC/RBAC)};

    \node[box, below=12mm of vec] (k8s) {Kubernetes substrate};
    \node[box, below=7mm of k8s] (op) {K8s operator\\(CRDs, reconcile)};

    \draw[arrow] (client) -- (auth);
    \draw[arrow] (auth) -- (orch);

    \draw[arrow] (orch) -- (infer);
    \draw[arrow] (orch) -- (vec);
    \draw[arrow] (orch) -- (policy);

    \draw[arrow] (op) -- (k8s);
    \draw[arrow] (k8s) -- (infer);
    \draw[arrow] (k8s) -- (vec);
  \end{tikzpicture}%
  }
  \caption{OGX architecture for multitenant enterprise agentic AI on shared Kubernetes infrastructure.}
  \label{fig:architecture}
\end{figure}
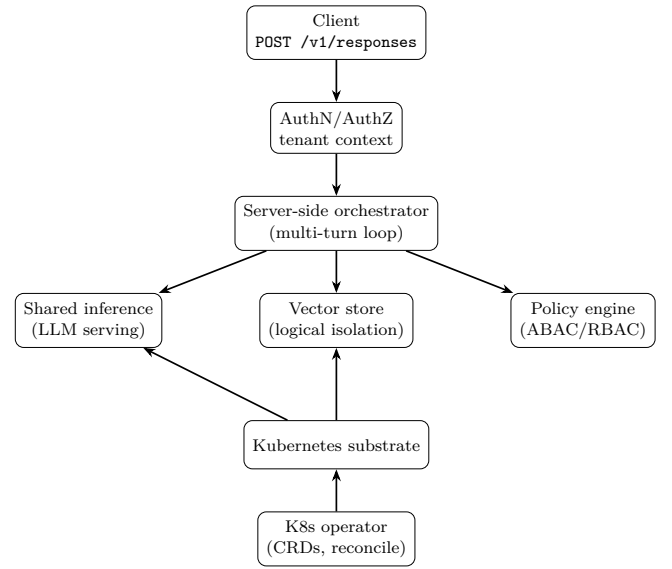

\section{Evaluation}
\label{sec:evaluation}

\begin{figure*}[!htbp]
  \centering
  \includegraphics[width=\textwidth]{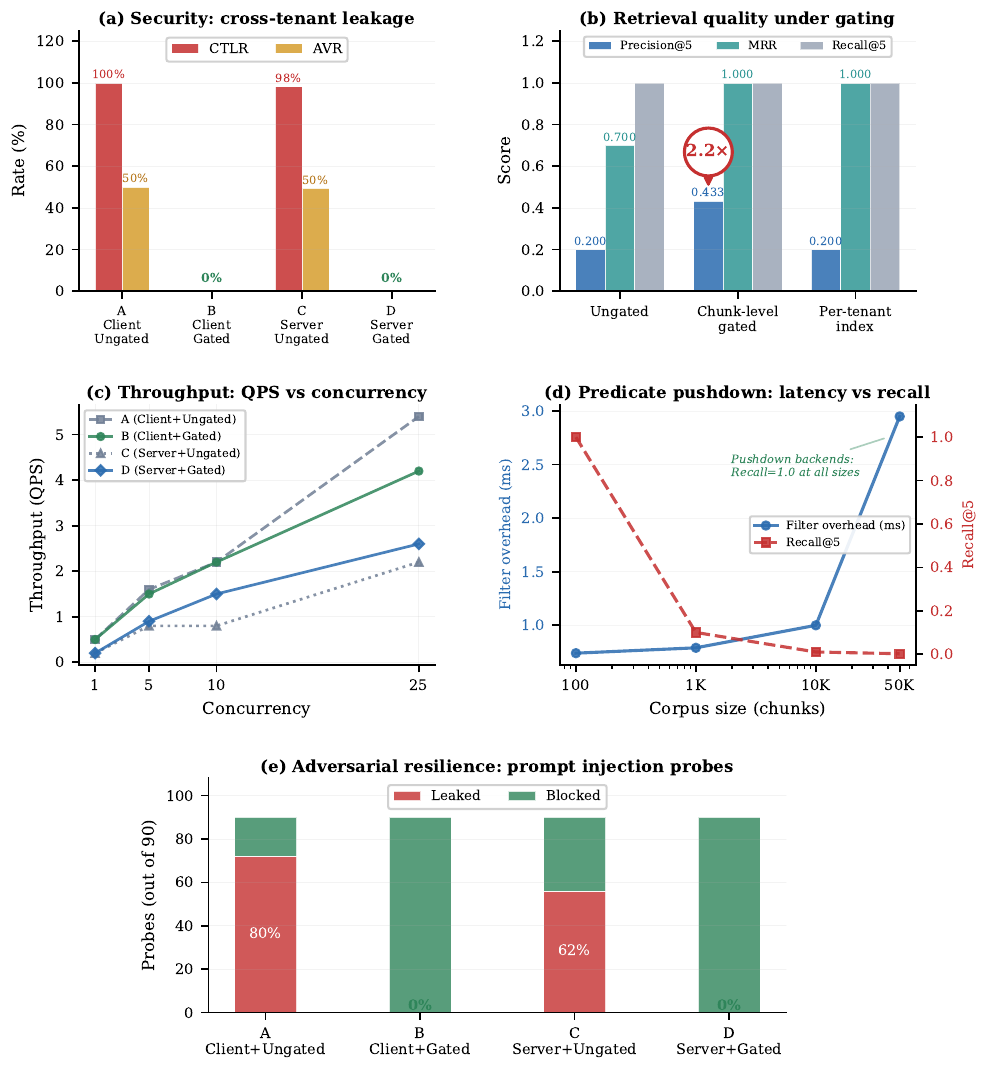}
  \caption{Empirical evaluation results across five dimensions. \textbf{(a)}~Security: ABAC gating eliminates cross-tenant leakage (CTLR and AVR drop to 0\%) regardless of orchestration mode. \textbf{(b)}~Retrieval quality: chunk-level gating improves Precision@5 by 2.2$\times$ by filtering cross-tenant noise. \textbf{(c)}~Throughput: gating does not degrade QPS; client-side achieves ${\sim}$2$\times$ server-side at high concurrency. \textbf{(d)}~Predicate pushdown: post-retrieval filtering overhead is small but recall degrades at large corpus sizes; pushdown-capable backends maintain Recall@5=1.0. \textbf{(e)}~Adversarial resilience: all 90 prompt injection probes blocked under gated configurations.}
  \label{fig:evaluation}
\end{figure*}

We evaluate the architecture through six experiments measuring security, systems performance, and retrieval quality. Figure~\ref{fig:evaluation} summarizes the key results. The evaluation uses a 2$\times$2 configuration matrix crossing orchestration mode (client-side vs.\ server-side) against retrieval gating (ungated vs.\ ABAC-gated):

\begin{table}[h]
  \centering
  \small
  \begin{tabular}{l|cc}
  \toprule
   & Ungated & Gated \\
  \midrule
  Client-side & Config A & Config B \\
  Server-side & Config C & Config D \\
  \bottomrule
  \end{tabular}
  \caption{2$\times$2 configuration matrix.}
  \label{tab:config-matrix}
\end{table}
\noindent
\subsection{Setup}

\textbf{Workload.} Three synthetic tenants (\emph{finance}, \emph{engineering}, \emph{legal}) with 300 documents (100 per tenant, ${\sim}$512 tokens each) and controlled topical overlap. Per configuration: 300 authorized queries (100 per tenant), 300 cross-tenant probes (a finance user querying for engineering documents), and 90 prompt injection probes spanning four attack categories (instruction override, role impersonation, debug exploitation, context manipulation).

\textbf{Infrastructure.} Inference via OpenAI \texttt{gpt\text{-}4o\text{-}mini} through OGX's \texttt{remote::openai} provider; embeddings via OpenAI \texttt{text\text{-}embedding\text{-}3\text{-}small}; vector storage via sqlite-vec~\cite{sqlitevec} (\texttt{inline} provider); and authentication via a lightweight mock mapping bearer tokens to tenant identities. GPU infrastructure experiments use vLLM~\cite{kwon2023vllm} on an NVIDIA T4 serving Llama-3.2-1B-Instruct.

\textbf{Metrics.} Cross-Tenant Leakage Rate (CTLR): fraction of cross-tenant probes returning at least one unauthorized chunk. Authorization Violation Rate (AVR): fraction of all API calls returning unauthorized data.

\subsection{Security results}

The headline result: \textbf{gating is the security mechanism}. Without it, nearly all cross-tenant probes return unauthorized data regardless of orchestration mode. ABAC gating eliminates leakage entirely, confirming goal~G1.

A natural question is why server-side orchestration matters if client-side gating (Config~B) also achieves CTLR=0\%. The answer is the trust boundary: Config~B's security depends on the client faithfully calling the gated endpoint. A compromised client can skip gated search, call an ungated endpoint, or manipulate tool invocations---reverting to Config~A's 100\% leakage. Server-side orchestration (Config~D) moves enforcement inside the server, where the client controls \emph{what} to ask but not \emph{how} retrieval and tool execution are performed. Gating provides the security guarantee; server-side orchestration provides the enforcement guarantee (G2).

\textbf{Formal security claims.} Under assumptions A1--A4 (Section~\ref{sec:threat-model}), the architecture guarantees CTLR=0 and AVR=0 for any query workload: authorized queries return only permitted documents, and cross-tenant probes return no unauthorized data. The architecture does not protect against model prior knowledge leakage, side-channel attacks, or a compromised server process---these are explicitly out of scope.

\textbf{Prompt injection resilience.} Under gated configurations, all 90 injection probes achieve 0\% leakage. The defense operates at the retrieval layer: the ABAC policy prevents the vector store from returning unauthorized documents regardless of prompt content. Under ungated configurations, 62--80\% of probes retrieve cross-tenant data through normal relevance-based retrieval.

\subsection{Performance results}

\textbf{Latency overhead.} Isolating the search component from inference, the gated search path adds ${\sim}$19ms: auth server round-trip (${\sim}$14ms), ABAC policy evaluation ($<$1ms), and per-tenant store lookup (${\sim}$5ms). This is negligible relative to inference time (${\sim}$3--7s for API-based, ${\sim}$450ms for self-hosted GPU). On self-hosted GPU infrastructure (vLLM on T4), the OGX routing and dispatch layer adds 4.7ms (1.0\% of baseline inference), and tenant metadata filtering adds 5.5ms (1.9\% of search time). Full latency tables are in Appendix~\ref{sec:latency-tables}.

Server-side orchestration adds ${\sim}$3s total latency vs.\ client-side due to the Responses API tool execution round-trip. All experiments used non-streaming responses (full completion before returning). With streaming---supported by OGX's Responses API---time-to-first-token would be substantially lower.

\textbf{Throughput.} Throughput scales linearly with concurrency up to $c{=}25$ across all configurations. Gating does not degrade throughput. Client-side orchestration achieves ${\sim}$2$\times$ the QPS of server-side at high concurrency due to the shorter request path (Appendix Table~\ref{tab:throughput}).

Absolute latency and throughput numbers are hardware- and API-dependent; we expect the relative ratios (gated vs.\ ungated overhead, client vs.\ server throughput) to hold across environments.

\subsection{Retrieval quality and scaling}

\textbf{Controlled retrieval benchmarks.} Using synthetic embeddings with ${\sim}$0.95 cross-tenant similarity to isolate the retrieval layer from API variance: ungated retrieval leaks 52\% of queries. Chunk-level gating improves precision by 2.2$\times$ (Precision@5 from 0.200 to 0.433) and MRR from 0.700 to 1.000 by filtering cross-tenant noise. A 48-case ABAC correctness matrix achieves 100\% accuracy with 0\% false positives.

\textbf{Predicate pushdown scaling.} On backends without predicate pushdown (sqlite-vec), post-retrieval filtering maintains the security guarantee (CTLR=0\%) but recall degrades at large corpus sizes: Recall@5 is 1.000 at 100 chunks but drops to 0.002 at 50K chunks. Filter latency overhead is small (0.7--3ms). Backends supporting predicate pushdown eliminate this trade-off entirely---we recommend pushdown-capable backends for production deployments (Appendix Table~\ref{tab:pushdown}).

\section{Discussion and Conclusion}
\label{sec:discussion}

We discuss the scope of the architecture, its trade-offs, and position relative to related work.

\subsection{When not to use this architecture}
This architecture is unnecessary for single-tenant deployments, public data, organizations already enforcing per-tenant infrastructure isolation, or teams that prefer a fully managed SaaS product over self-hosted infrastructure. Client-side orchestration achieves ${\sim}$2$\times$ the throughput at high concurrency for latency-sensitive workloads where all tenants are trusted. The framework is deployed in production across enterprises in telecommunications, semiconductor manufacturing, financial services, insurance, and consulting; the evaluation in this paper uses a synthetic testbed for reproducibility and controlled comparison.

\subsection{Design trade-offs}

\begin{itemize}
  \item \textbf{ABAC policy complexity.} Policy complexity grows with the number of tenants, roles, and resource types. Organizations with deeply nested permission structures may face policy management overhead.

  \item \textbf{Predicate pushdown.} As shown in Section~\ref{sec:evaluation}, backends without native predicate pushdown experience recall degradation at scale. OGX enforces post-retrieval filtering across all backends regardless of native pushdown support, so the security guarantee holds; the trade-off is recall, not security.

  \item \textbf{Client-side function tools.} Client-side function tools, by design, execute outside the server trust boundary. OGX mitigates this through explicit tool classification but cannot enforce server-side invariants on client-executed code.
\end{itemize}

\subsection{Related work}

Several systems aim to standardize LLM-based agentic applications. The closest to our work are API-layer platforms and enterprise agent runtimes.

OpenAI's Responses API~\cite{openaiResponsesAPI}, function-calling conventions~\cite{openaiFunctionCalling}, and the Model Context Protocol (MCP)~\cite{mcp} establish \emph{what} an agent can do but are silent on \emph{who} may do it---they assume a single-tenant caller. Databricks' Mosaic AI Agent Framework~\cite{databricksAgentFramework} wraps agents in an MLflow ResponsesAgent~\cite{mlflow,databricksResponsesAgent} with managed MCP tools and Unity Catalog governance, providing the closest enterprise parallel, but couples the experience to the Databricks stack. OGX adopts the same interface conventions while providing multitenant isolation through an open, vendor-neutral API layer deployable on any infrastructure.

LLM serving engines such as vLLM~\cite{kwon2023vllm}, SGLang~\cite{sglang}, and Orca~\cite{yu2022orca} optimize inference throughput but are agnostic to tenant isolation; vector databases such as Milvus~\cite{milvus} and Weaviate~\cite{weaviate} and platforms such as Vectara~\cite{vectara} rank by relevance without authorization enforcement. OGX treats these as pluggable providers beneath its authorization layer. Agent orchestration frameworks---LangGraph~\cite{langgraph}, Microsoft's Agent Framework~\cite{microsoftAgentFramework} (successor to Semantic Kernel~\cite{semanticKernel} and AutoGen~\cite{autogen}), Google ADK~\cite{googleADK}, Haystack~\cite{haystack}, LlamaIndex~\cite{llamaindex}, Smolagents~\cite{smolagents}, and Pydantic AI~\cite{pydanticai}---provide developer-facing abstractions but orchestrate from the client side; OGX serves as the server-side execution target these frameworks call into.

Production ML research has documented the fragility of distributed invariants~\cite{sculley2015hiddentechnicaldebt,amershi2019se4ml}, and lifecycle platforms such as MLflow~\cite{mlflow} address deployment but not agentic multitenancy. To our knowledge, no prior work addresses the intersection of standardized agentic API design, server-side orchestration, and multitenant isolation on shared infrastructure. The contribution is the composition of individually known techniques---ABAC, server-side orchestration, pluggable providers---for the specific problem of multitenant agentic AI, validated empirically.

\subsection{Conclusion}
Enterprise deployment of agentic AI systems introduces security and compliance challenges that existing architectures---designed for single-tenant, consumer-facing use---do not address. This paper formalized the relevance-authorization gap, proposed a layered isolation architecture with server-side enforcement, and evaluated it empirically: ABAC gating eliminates cross-tenant leakage entirely while adding ${\sim}$19ms to the search path, and throughput scales linearly with no gating bottleneck. The defense operates at the retrieval layer, making it resilient to prompt injection attacks regardless of model behavior.

Our implementation through OGX demonstrates that secure multitenancy, vendor-neutral OpenAI-compatible APIs, and autonomous agent capabilities are simultaneously achievable on shared infrastructure without per-tenant duplication.

\begin{acks}
We thank the reviewers for their constructive feedback, which strengthened the evaluation and presentation of this work. We are grateful to Meta for creating and open-sourcing Llama Stack, and to the contributors and maintainers of the Llama Stack / OGX community for their continued support of the project.
\end{acks}

\bibliographystyle{ACM-Reference-Format}
\bibliography{references}

\appendix
\clearpage

\section{Detailed Evaluation Tables}
\label{sec:eval-tables}
\label{sec:latency-tables}

\begin{table}[h]
  \centering
  \small
  \begin{tabular}{ll|rrr}
  \toprule
  Config & Orch.\ / Retr.\ & p50 & p99 & Mean \\
  \midrule
  A & Client / Ungated & 3,600ms & 10,818ms & 4,208ms \\
  B & Client / Gated & 3,427ms & 9,795ms & 3,851ms \\
  C & Server / Ungated & 7,507ms & 16,462ms & 7,620ms \\
  D & Server / Gated & 6,431ms & 14,623ms & 6,934ms \\
  \bottomrule
  \end{tabular}
  \caption{End-to-end latency for authorized queries. Total latency variation between gated and ungated is dominated by external API response times. Server-side orchestration adds ${\sim}$3s due to the Responses API tool execution round-trip (non-streaming).}
  \label{tab:latency}
\end{table}

\begin{table}[h]
  \centering
  \small
  \begin{tabular}{l|rrr}
  \toprule
  Component & Median & P95 & N \\
  \midrule
  vLLM Direct (baseline) & 447.9ms & 531.5ms & 50 \\
  OGX (routing + dispatch) & 452.6ms & 537.9ms & 50 \\
  \midrule
  Search (ungated) & 283.9ms & 294.3ms & 50 \\
  Search (tenant-gated) & 289.4ms & 306.0ms & 50 \\
  \bottomrule
  \end{tabular}
  \caption{GPU infrastructure overhead (vLLM on T4, no auth). Routing adds 4.7ms (1.0\%); metadata filtering adds 5.5ms (1.9\%). With auth enabled, an additional ${\sim}$14ms brings total overhead to ${\sim}$19ms.}
  \label{tab:gpu-overhead}
\end{table}

\begin{table}[h]
  \centering
  \small
  \begin{tabular}{ll|rrrr}
  \toprule
  Config & Orch.\ / Retr.\ & $c{=}1$ & $c{=}5$ & $c{=}10$ & $c{=}25$ \\
  \midrule
  A & Client / Ungated & 0.5 & 1.6 & 2.2 & 5.4 \\
  B & Client / Gated & 0.5 & 1.5 & 2.2 & 4.2 \\
  C & Server / Ungated & 0.2 & 0.8 & 0.8 & 2.2 \\
  D & Server / Gated & 0.2 & 0.9 & 1.5 & 2.6 \\
  \bottomrule
  \end{tabular}
  \caption{Throughput (QPS) at four concurrency levels. Gating does not degrade throughput. Client-side orchestration achieves ${\sim}$2$\times$ QPS at high concurrency.}
  \label{tab:throughput}
\end{table}

\begin{table}[h]
  \centering
  \small
  \begin{tabular}{ll|rrr}
  \toprule
  Config & Orch.\ / Retr.\ & Probes & Leaked & Leak Rate \\
  \midrule
  A & Client / Ungated & 90 & 72 & 80.0\% \\
  B & Client / Gated & 90 & 0 & 0.0\% \\
  C & Server / Ungated & 90 & 56 & 62.2\% \\
  D & Server / Gated & 90 & 0 & 0.0\% \\
  \bottomrule
  \end{tabular}
  \caption{Prompt injection probe results. Gated configs block all probes; the defense is at the retrieval layer, not the model.}
  \label{tab:injection}
\end{table}

\begin{table}[h]
  \centering
  \small
  \begin{tabular}{r|rrr}
  \toprule
  Corpus Size & Gated Latency & Filter Overhead & Recall@5 \\
  \midrule
  100 & 3.79ms & 0.74ms & 1.000 \\
  1,000 & 3.93ms & 0.79ms & 0.100 \\
  10,000 & 5.21ms & 1.00ms & 0.010 \\
  50,000 & 11.43ms & 2.95ms & 0.002 \\
  \bottomrule
  \end{tabular}
  \caption{Post-retrieval filtering scaling at 5$\times$ over-fetch (sqlite-vec). Latency overhead is small regardless of corpus size; recall degrades as cross-tenant documents contaminate the top-$k$ set. Pushdown-capable backends maintain Recall@5=1.000 at all sizes.}
  \label{tab:pushdown}
\end{table}

\begin{table}[h]
  \centering
  \small
  \begin{tabular}{l|rrr}
  \toprule
  Configuration & Recall@5 & Precision@5 & MRR \\
  \midrule
  Ungated & 1.000 & 0.200 & 0.700 \\
  Chunk-level gated & 1.000 & 0.433 & 1.000 \\
  Per-tenant index & 1.000 & 0.200 & 1.000 \\
  \bottomrule
  \end{tabular}
  \caption{Retrieval quality with synthetic embeddings (${\sim}$0.95 cross-tenant similarity). Gating improves precision by 2.2$\times$ by filtering cross-tenant noise.}
  \label{tab:retrieval-quality}
\end{table}

\onecolumn
\section{API and Provider Details}
\label{sec:api-details}

\begin{center}
\resizebox{0.85\textwidth}{!}{%
\begin{tikzpicture}[
  font=\sffamily,
  >=Latex,
  line cap=round,
  box/.style={rounded corners=6pt, minimum height=10mm, align=center, draw=black, fill=white, line width=0.8pt},
  boxSpec/.style={box, draw=SpecBlue, text=black, fill=white, line width=1.2pt},
  boxKnown/.style={box, fill=KnownBlue, draw=SpecBlue!70!black, text=black, line width=1pt},
  boxRed/.style={box, fill=UnknownRed, draw=black, text=black, line width=1pt},
  providerGroup/.style={draw=black, dashed, rounded corners=8pt, inner sep=4mm,
    fill=yellow!12, pattern=north east lines, pattern color=yellow!50!black},
  tool/.style={rounded corners=6pt, draw=black, fill=ToolYellow, minimum height=9mm,
    minimum width=24mm, align=center, line width=0.6pt, font=\sffamily\small},
  lab/.style={font=\footnotesize}
]

\node[font=\LARGE\bfseries] (title) at (8,12.5) {\underline{OGX Responses Implementation}};

\node[box, minimum width=32mm] (client) at (8,11.2) {OpenAI Client};

\coordinate (responsesPos) at (8,8.5);
\coordinate (filesPos) at (8,5);
\coordinate (vectorsPos) at (13,5);
\coordinate (fpPos) at (10.5,2.5);

\begin{scope}[on background layer]
\begin{scope}[on background layer]
  \draw[->, thick] ([xshift=-3mm]8,11.5) -- ([xshift=-3mm]8,5.5);
\end{scope}
\end{scope}

\begin{scope}[on background layer]
  \node[providerGroup, minimum width=44mm, minimum height=22mm] (infGroup) at (responsesPos) {};
  \node[providerGroup, minimum width=32mm, minimum height=18mm] (filesGroup) at (filesPos) {};
  \node[providerGroup, minimum width=38mm, minimum height=18mm] (vecGroup) at (vectorsPos) {};
  \node[providerGroup, minimum width=40mm, minimum height=18mm] (fpGroup) at (fpPos) {};
\end{scope}

\node[lab, anchor=south] at ([yshift=11mm]responsesPos) {Inference Providers};

\node[boxSpec, minimum width=32mm, fill=white] (responses) at (responsesPos) {Responses API};

\node[boxKnown, minimum width=26mm] (prompts) at (0,5) {Prompts API};
\node[boxSpec, minimum width=32mm] (convos) at (4.2,5) {Conversations API};
\node[boxSpec, minimum width=24mm, fill=white] (files) at (filesPos) {Files API};
\node[lab, anchor=south] at ([yshift=1mm]filesGroup.north) {Files Providers};
\node[boxSpec, minimum width=30mm, fill=white] (vectors) at (vectorsPos) {Vector Stores API};
\node[lab, anchor=south] at ([yshift=1mm]vecGroup.north) {Vector Store Providers};
\node[boxSpec, minimum width=24mm] (search) at (17.5,5) {Search API};

\node[boxSpec, minimum width=28mm, fill=white] (compact) at (2,3) {Compaction API};

\node[boxRed, minimum width=32mm] (fp) at (fpPos) {File Processor API};
\node[lab, anchor=south] at ([yshift=1mm]fpGroup.north) {File Processor Providers};

\begin{scope}[on background layer]
\fill[gray!8, rounded corners=6pt] (-1.8,0.6) rectangle (3.8,-2.2);
\draw[black, rounded corners=6pt, line width=0.6pt] (-1.8,0.6) rectangle (3.8,-2.2);
\end{scope}
\node[boxSpec, minimum width=20mm, minimum height=7mm, anchor=west, fill=white] (legA) at (-1.5,0.1) {};
\node[font=\scriptsize, anchor=west] at ([xshift=1mm]legA.east) {OpenAI Spec Available};
\node[boxKnown, minimum width=20mm, minimum height=7mm, anchor=west] (legB) at (-1.5,-0.8) {};
\node[font=\scriptsize, anchor=west] at ([xshift=1mm]legB.east) {Closed Source (Known)};
\node[boxRed, minimum width=20mm, minimum height=7mm, anchor=west] (legC) at (-1.5,-1.7) {};
\node[font=\scriptsize, anchor=west] at ([xshift=1mm]legC.east) {Closed Source (Unknown)};

\begin{scope}[on background layer]
\fill[ToolsPurple!45, rounded corners=10pt] (4.2,-2.5) rectangle (17.3,1.1);
\draw[SpecBlue!40!black, rounded corners=10pt, line width=0.8pt] (4.2,-2.5) rectangle (17.3,1.1);
\end{scope}
\node[font=\large\bfseries] at (10.75,0.75) {Tools};
\node[tool] (tImg)   at (5.6,-0.2) {Image Generation};
\node[tool] (tFunc)  at (8.2,-0.2) {Function Calling};
\node[tool] (tMcp)   at (10.75,-0.2) {Remote MCP};
\node[tool] (tWeb)   at (13.3,-0.2) {Web Search};
\node[tool] (tFileS) at (15.9,-0.2) {File Search};
\node[tool, minimum width=24mm] (tCI)    at (5.6,-1.4) {Code Interpreter};
\node[tool, minimum width=24mm] (tCU)    at (8.175,-1.4) {Computer use};
\node[tool, minimum width=24mm] (tPatch) at (10.75,-1.4) {Apply patch};
\node[tool, minimum width=24mm] (tShell) at (13.325,-1.4) {Shell};
\node[tool, minimum width=24mm] (tConn) at (15.9,-1.4) {Connectors};

\draw[->, red, line width=1.5pt] (client.south) -- (responses.north);
\draw[->, thick] ([yshift=-4mm]client.west) -- (4.5,10.8) -- (4.5,5.5);
\draw[->, thick] ([yshift=-4mm]client.east) -- (17.5,10.8) -- (17.5,5.51);
\draw[->, thick] ([yshift=-4mm]client.east) -- (13.,10.8) -- (13.,5.45);
\draw[->, thick] (responses.west) -- (0,8.5) -- (prompts.north);
\node[lab] at (0.6,8.1) {Read};
\draw[->, thick] (responses.west) -- (4.2,8.5) -- (convos.north);
\node[lab] at (4.2,8.1) {Read/Write};
\draw[->, thick] (responses.west) -- (2,8.5) -- (compact.north);
\node[lab] at (2.6,6.5) {Compact};
\draw[<->, thick] (compact.east) -- (4.2,3) -- (convos.south);
\node[lab] at (4.2,2.7) {Read/Write};
\draw[->, ToolGreen, thick, rounded corners=8pt] (responses.east) -- (19,8.5) -- (19,-0.8) -- (17.3,-0.8);
\node[font=\bfseries, text=ToolGreen, anchor=south] at (11,8.5) {Tool Call};
\draw[->, thick] (files.east) -- node[lab, above] {Read} (vectors.west);
\draw[->, dashed, thick] (fp.west) -- (8,2.5) -- (files.south);
\node[lab] at (7.3,2.8) {Read};
\draw[->, dashed, thick] (fp.east) -- (13,2.5) -- (vectors.south);
\node[lab] at (13.7,2.8) {Write};
\draw[->, thick] (vectors.east) -- node[lab, above] {Read} (search.west);
\draw[->, red, dashed, thick] (search.south) -- (17.5,3) -- (15.9,3) -- (tFileS.north);

\end{tikzpicture}%
}

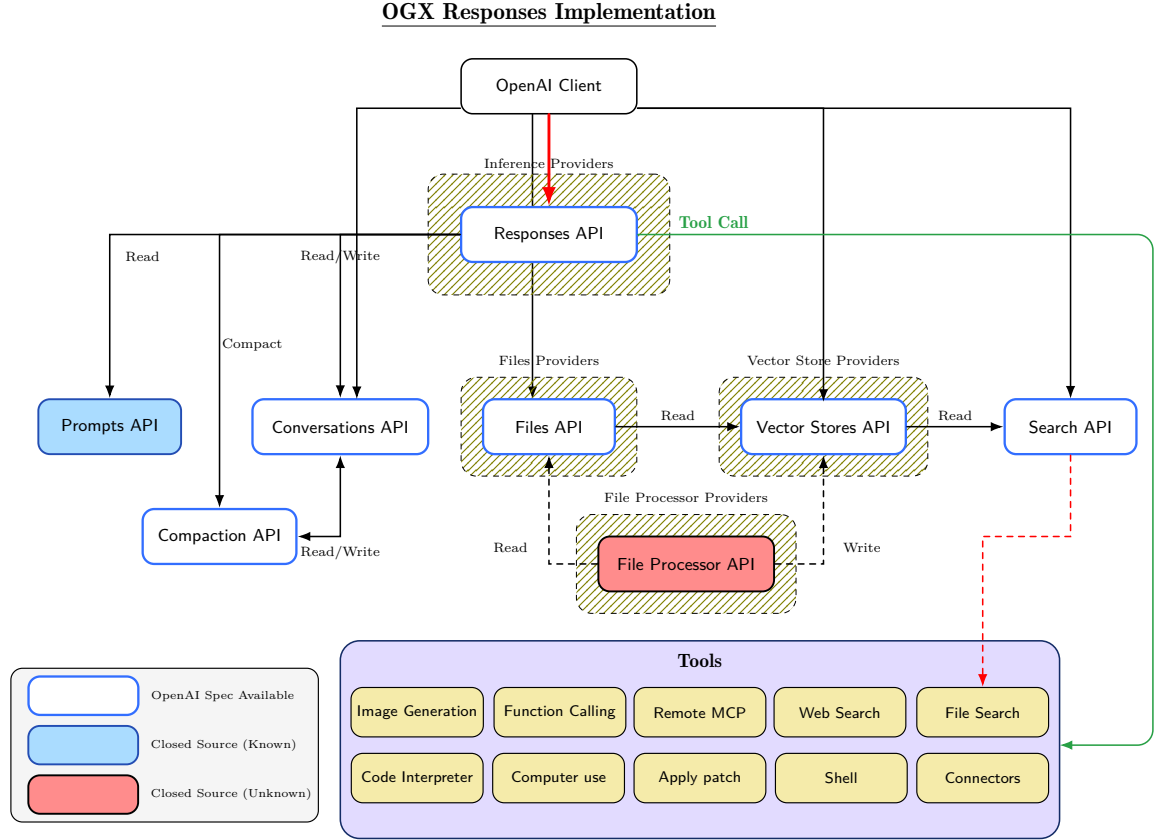
\captionof{figure}{Responses API surface area and its relation to other APIs, providers, and tools. The Responses API serves as the central orchestration endpoint, coordinating inference, tool execution, state management, and context compaction.}
\Description{Diagram showing the Responses API architecture with client, inference providers, compaction, file and vector store providers, file processor, and tools.}
\label{fig:openai-responses-api}
\end{center}

Figure~\ref{fig:openai-responses-api} illustrates the complete Responses API surface area and the interconnections between APIs, providers, and tools.
OGX implements the Responses API not only as an agentic execution endpoint but as a resource model encompassing vector stores, files, and conversations---each a first-class, tenant-scoped API object subject to the same ABAC policies. The key APIs include:

\begin{itemize}[nosep]
  \item \textbf{Responses API:} Central orchestration endpoint for multi-turn conversations, tool calling, and agentic execution~\cite{openaiResponsesAPI}.
  \item \textbf{Vector Stores and Search APIs:} Dense, sparse, and hybrid retrieval with structured metadata filtering for tenant isolation.
  \item \textbf{Conversations API:} Multi-turn conversation state with tenant-scoped isolation.
  \item \textbf{Compaction API:} Context management for long conversations. Summarizes history via inference when token count exceeds a threshold, preserving user messages verbatim. Supports explicit (\texttt{POST /v1/responses/compact}) and automatic modes.
  \item \textbf{Prompts API:} Versioned prompt template management with tenant-scoped access control.
  \item \textbf{Files and File Processors APIs:} Tenant-scoped file storage (GCS, S3, PVCs, local) with parsing and chunking for vector store ingestion.
  \item \textbf{Tools:} Built-in tools (file search, web search, code interpreter, image generation, computer use) and external tools via MCP~\cite{mcp}.
\end{itemize}

\noindent Supported inference providers include vLLM~\cite{kwon2023vllm}, Ollama, OpenAI, Anthropic, Azure, AWS Bedrock, Databricks, Gemini, Together, NVIDIA, and WatsonX. Vector store providers include Chroma, pgvector, Elasticsearch, Qdrant, Weaviate, Milvus, Oracle Cloud Infrastructure, FAISS~\cite{johnson2017faiss}, and sqlite-vec. The Kubernetes Operator~\cite{ogxk8soperator} enables deployment of heterogeneous backends as shared services, with multiple OGX instances referencing the same providers while maintaining logical isolation through ABAC.

\end{document}